\documentclass[sigconf]{acmart}

\usepackage{booktabs} 

\usepackage[ruled]{algorithm2e} 
\usepackage{hyperref}
\usepackage{url}
\usepackage{graphicx}
\graphicspath{ {./} }
\usepackage{mathtools}
\usepackage{amsmath}
\usepackage{graphicx}
\usepackage{subcaption}

\setcopyright{rightsretained}

\acmConference[]{}{}{}
\acmYear{2018}
\copyrightyear{2018}

\acmArticle{4}
\acmPrice{15.00}

\begin{document}
\title{Deep density networks and uncertainty in recommender systems}

\author{Yoel Zeldes}
\affiliation{%
  \institution{Taboola}
  \city{Tel Aviv}
  \state{Israel}
}
\email{yoel.z@taboola.com}

\author{Stavros Theodorakis}
\affiliation{%
  \institution{DeepLab}
  \city{Athens}
  \state{Greece}
}
\email{sth@deeplab.ai}

\author{Efrat Solodnik}
\affiliation{%
  \institution{Taboola}
  \city{Tel Aviv}
  \state{Israel}
}
\email{efrat.s@taboola.com}

\author{Aviv Rotman}
\affiliation{%
  \institution{Taboola}
  \city{Tel Aviv}
  \state{Israel}
}
\email{aviv.r@taboola.com}

\author{Gil Chamiel}
\affiliation{%
  \institution{Taboola}
  \city{Tel Aviv}
  \state{Israel}
}
\email{gil.c@taboola.com}

\author{Dan Friedman}
\affiliation{%
  \institution{Taboola}
  \city{Tel Aviv}
  \state{Israel}
}
\email{dan.f@taboola.com}

\renewcommand{\shortauthors}{B. Trovato et al.}

\newcommand{\fix}{\marginpar{FIX}}
\newcommand{\new}{\marginpar{NEW}}
\newcommand{\indep}{\rotatebox[origin=c]{90}{$\models$}}
\renewcommand\shortauthors{Zeldes, Y. et al}
\newcommand{\matr}[1]{\mathbf{#1}}
\newcommand{\vect}[1]{\mathbf{#1}}

\begin{abstract}
Building robust online content recommendation systems requires learning complex interactions between user preferences and content features. The field has evolved rapidly in recent years from traditional multi-arm bandit and collaborative filtering techniques, with new methods employing Deep Learning models to capture non-linearities. Despite progress, the dynamic nature of online recommendations still poses great challenges, such as finding the delicate balance between exploration and exploitation.  
In this paper we show how uncertainty estimations can be incorporated  
by employing them in an optimistic exploitation/exploration strategy for more efficient exploration of new recommendations. 
We provide a novel hybrid deep neural network model, Deep Density Networks (DDN), which integrates content-based deep learning models with a collaborative scheme that is able to robustly model and estimate uncertainty.  
Finally, we present online and offline results after incorporating DNN into a real world content recommendation system that serves billions of recommendations per day, and show the benefit of using DDN in practice. 
\end{abstract}

%
%
%


\maketitle

\section{Introduction}

\begin{figure*}[t]
\centering
\includegraphics[scale=0.5]{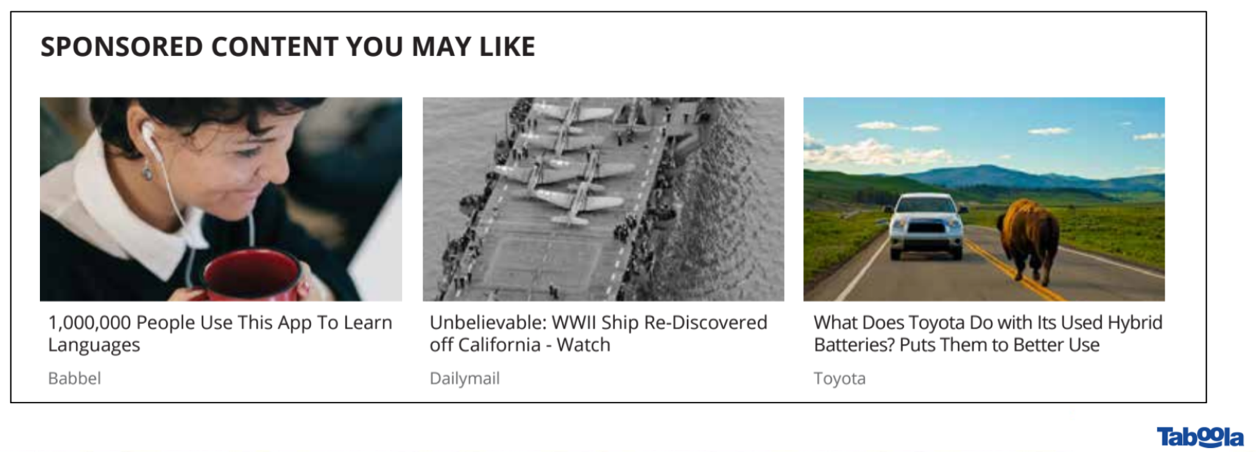}
\caption{Taboola's recommendation widget example.}
\label{fig:Widget}
\end{figure*}

In order to navigate the vast amounts of content on the internet, users either rely on search queries, or on content recommendations powered by algorithms. 
Taboola's content discovery platform  
leverages computational models to match content to users who are likely to engage with it. Taboola's content recommendations are shown in widgets that are usually placed at the bottom of articles (see Fig.~\ref{fig:Widget}) in various websites across the internet, and serve billions of recommendations per day, with a user base of hundreds of millions of active users.

Modeling in recommendation systems can be classified into either Collaborative Filtering (CF) or content-based methods. 
CF methods use past user-item interactions to predict future ratings~\cite{linden2003amazon} 
usually realized by Matrix Factorization (MF)~\cite{mnih2008probabilistic}. 
A drawback to MF approaches is the cold-start (CS) problem. 
Content-based approaches mitigate CS by modeling explicitly meta-information about the items. 
This can be seen as a trade-off between memorization of users/items seen in the past, and generalization for new items. 
Hybrid methods that combine both the memorization and generalization advantages 
have also been proposed~\cite{cheng2016wide}. 
We use this kind of hybrid approach, by employing deep neural networks (DNNs) to learn item representations and combining those with contextual features.

In order to improve long-term performance and tackle faster the CS problem, recommender systems have been modeled in a multi-arm bandit setting, where the goal is to find an \textit{exploitation} and \textit{exploration} selection strategy that maximizes the long term reward~\cite{li2010contextual}. 
One of the basic approaches to deal with multi-arm bandit problems is the $\epsilon$-greedy algorithm. 
Upper Confidence Bound (UCB)~\cite{auer2002finite} and Thompson sampling techniques~\cite{thompson1933likelihood} use uncertainty estimations in order to perform more efficient exploration of the feature space, either by explicitly adding the uncertainty to the estimation or by sampling from the posterior distribution respectively. 
Estimating uncertainty is crucial in order to utilize these methods.
%
%
To deal with this, bayesian neural networks~\cite{neal2012bayesian} using distributions over the weights were applied by using either sampling or stochastic variational inference~\cite{kingma2013auto, rezende2014stochastic}. 
%
\cite{blundell2015weight} proposed Bayes by Backprop algorithm for the variational posterior estimation and applied Thompson sampling in a multi-arm bandit setting similarly to our case. 
\cite{gal2016dropout} proposed Monte Carlo (MC) dropout, a Bayesian approximation of model uncertainty achieved 
by extracting estimations from the different sub-models that have been trained 
using dropout.  
Building upon their previous work, the authors separated uncertainty into two types, model and data uncertainty, while studying the effect of each uncertainty separately in computer vision tasks~\cite{kendall2017uncertainties}. 
Similarly, we separate recommendation prediction uncertainty into three types: measurement, data and model uncertainty.  
In contrast to~\cite{zhu2017deep}, we assumed heteroscedastic data uncertainty which was a more natural choice for recommendation systems.  
%
%
%
Our work has parallels to~\cite{li2010contextual} where the authors formulated the exploration/exploitation trade-off in personalized article recommendation as a contextual bandit problem proposing LinUCB which adapts the UCB strategy. 
Our approach extends LinUCB by using a deep model instead, while explicitly modeling and estimating the different types of uncertainty. 
%

Finally, we model measurement noise using a Gaussian model and combine it with a Gaussian Mixture Model (GMM) to form a deep Mixture density network (MDN)~\cite{bishop1994mixture}. The effect of measurement noise and noisy labels has been studied extensively \cite{frenay2014classification}. We were inspired by 
\cite{mnih2012learning, goldberger2016training} where the authors in the former proposed a probabilistic model for the conditional probability of seeing a wrong label  
and in the latter explicitly modeled noise via a softmax layer.  

In this paper we introduce a unified hybrid DNN to explicitly model and estimate measurement, data and model uncertainty and utilize them to form an optimistic exploitation/exploration selection strategy that is applied in a real world and large-scale content recommendation system. 
We explicitly model recommendations' content and combine it with context by using a collaborative fusion scheme.  
To the best of our knowledge this is the first time that a hybrid DNN model with uncertainty estimations is employed in a multi-arm bandit setting for recommender systems.



\section{Taboola's recommender system overview}


\begin{figure*}[t]
\centering
\includegraphics[scale=0.3]{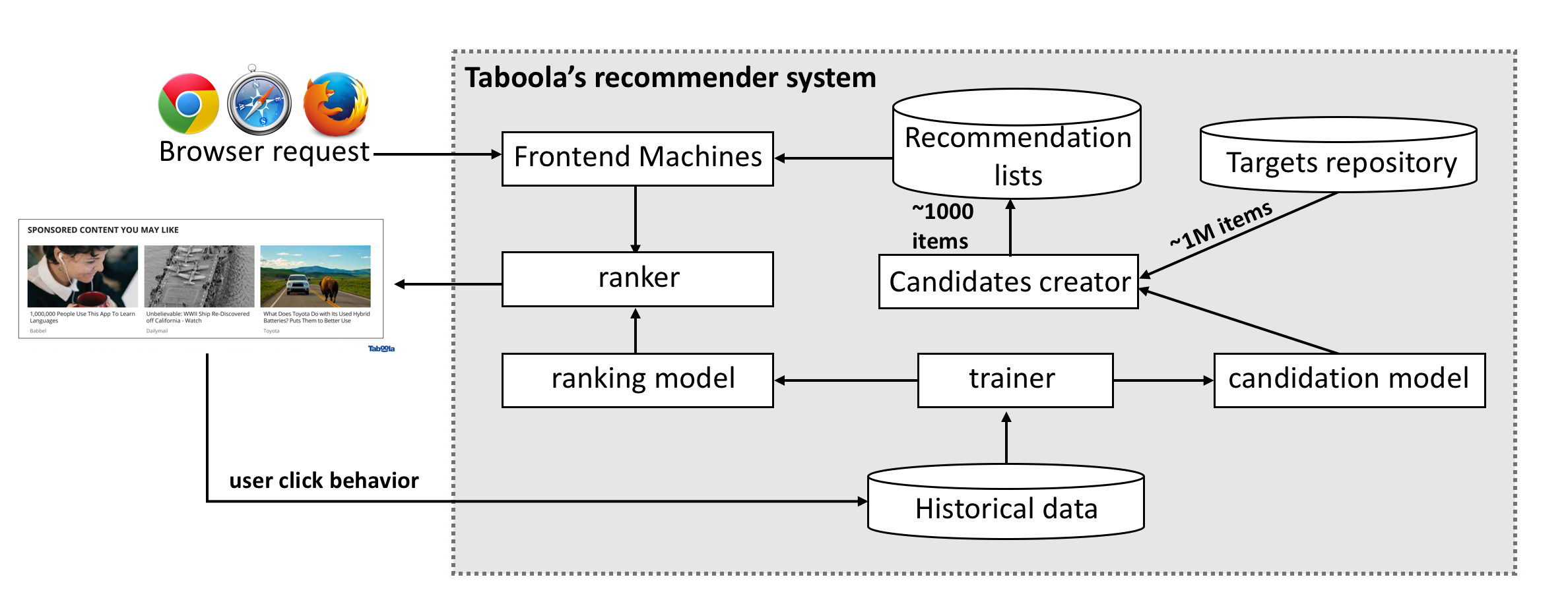}
\caption{High level overview of candidation and ranking architecture.}
\label{fig:Overview}
\end{figure*}

Taboola's revenue stream is facilitated by online advertisers, who pay a fixed amount \emph{CPC} 
(\textit{Cost Per Click}) 
for each click event on a Taboola recommendation. 
The algorithm's total value is measured in \emph{RPM} 
(\textit{Revenue Per Mille}) 
where \textit{$RPM = CTR*CPC*1000$} and \emph{CTR} 
is the average revenue accrued after showing a recommendation 1000 times, and \emph{CTR} (\textit{Click Through Rate}) 
is the click probability of a recommendation. 
Taboola's main algorithmic challenge is to provide an estimate of the CTR in any given context. 
Taboola's recommendation engine needs to provide recommendations within strict time constraints ($<50 ms$). 
As It is infeasable to rank millions of recommendations in that time frame,
in order to support this 
we have partitioned the system into
\textit{candidation} and \textit{ranking}~Fig.~\ref{fig:Overview}. 
During the candidation, we narrow down the list of possible recommendations 
based on features such as the visual appearance of the item and empirical click statistics. 
This relatively small list of recommendations is written to distributed databases in worldwide data centers, and are re-calculated by Taboola's servers continuously throughout the day. 
When we get request for recommendations, they retrieve the relevant ready-made recommendation list, and perform an additional ranking of the recommendations based on additional user features using a DNN, further personalizing recommendations. 
This system architecture shows similarities to (\cite{cheng2016wide}).

Due to the dynamic nature of Taboola's marketplace our algorithm needs to evaluate new recommendations, with tens of thousands of new possible recommendations every day. To support this, we split the algorithm into \textit{exploration} and \textit{exploitation} modules. Exploitation aims to choose the recommendations maximizing RPM, while exploration aims to enrich the dataset. 
%
In this paper we focus on the candidation phase and the corresponding CTR prediction task, leaving out of the scope the second ranking step.

\label{sec:overview}

\section{Deep density network}
\label{sec:mdn}
%

\begin{figure*}[t]
\centering
\includegraphics[scale=0.3]{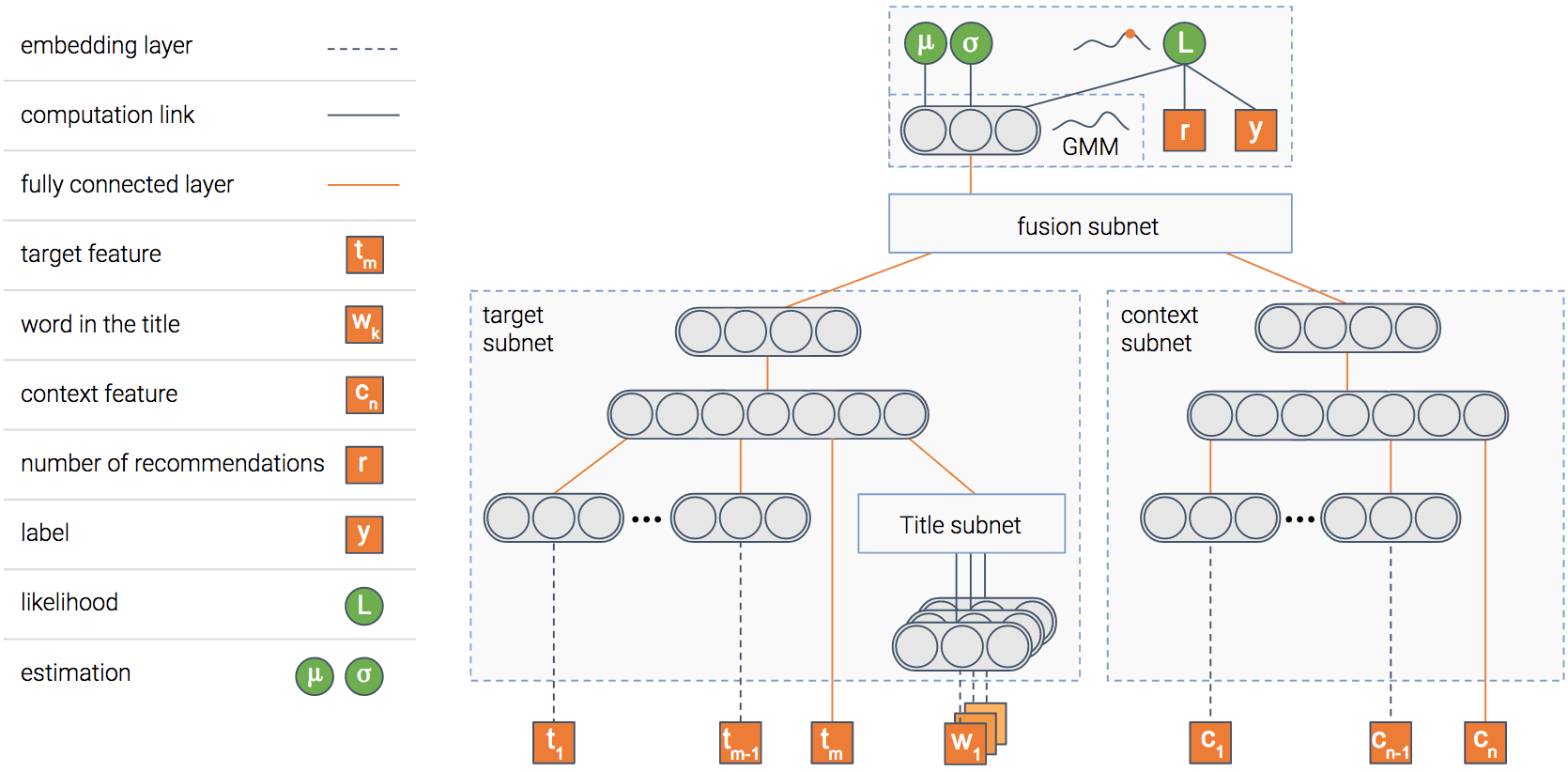}
\caption{A hybrid content-based and collaborative filtering model accounting for data and measurement uncertainties. Both target and context features are passed through a DNN, then through a fusion sub-network, and finally through a fully-connected layer that outputs the parameters of a GMM. The number of recommendations $r$ is used for attenuating the loss of noisy examples.}
\label{fig:model_architecture}
\end{figure*}

Our deep recommender model is a hybrid content-based and collaborative filtering (CF) system (Fig.~\ref{fig:model_architecture}). 
%
We use two DNN subnets to model target and context features. 
The target subnet gets as input the content features seen by user together with additional categorical features which are unseen to the user. 
The categorical features are passed through an embedding layer and 
concatenated with the content features, 
followed by fully-connected layers with a RELU activation function, resulting in the target feature descriptor. 
Similarly, the context features are modeled using a DNN, taking as input context features such as device type where the target is recommended, resulting in the context feature descriptor. 
The target and context feature descriptors are then fused in a collaborative filtering manner 
and finally passed through a fully-connected layer which outputs the parameters of  a GMM i.e. ($\alpha_i$, $\mu_i$ and $\sigma_i$) to form a MDN. This GMM model is employed in order to model  data uncertainty as discussed in sec.~\ref{sec:data_uncertainty}

In order to train our models, we use historical data which consists of target and context pairs $(t,c)$, where $t$ is the target we recommended in a specific browsing context $c$ accompanied with a binary variable which indicates if the recommendation was clicked by the user. 
A natural choice would be to estimate CTR using a logistic loss. 
However,  our data contains great variability in terms of CTR due to various factors which are external to the content itself; 
As an example, a widget which contains very large images will be more likely to capture the user's attention, which subsequently increases the probability of a click. 
Moreover, even inside a certain widget, the specific location of a recommendation (top left, bottom right) can have a vast impact on the eventual CTR, which is independent of the content itself. 
To account for this, building upon our previous work~\cite{TabPatent}, we use a calibrated version of the CTR, to diminish the variability due to different contexts in which a recommendation was shown. 
%
In practice we train our DNN network to predict the log of the calibrated CTR using Maximum Likelihood Estimation (MLE), as this  allow us to estimate unconstrained scalar values roughly normally distributed with zero-mean. 
From hereafter we will refer to log calibrated CTR simply as CTR. 

\section{Uncertainty in Recommender systems}
\label{sec:uncertainty}

We separate uncertainty into three different types: data, measurement, and model uncertainties, and study the role of each one in recommender systems.  In addition, we provide a deep unified framework for explicitly modeling and estimating all types and further exploit them to form an optimistic exploration strategy (sec.~\ref{sec:optimistic_stragegy}).

\subsection{Data Uncertainty}
\label{sec:data_uncertainty}

Data uncertainty corresponds to the inherent noise of the observations; it cannot be reduced even if more data was to be collected and is categorized into homoscedastic and heteroscedastic. 
Homoscedastic is constant over all different inputs, 
and heteroscedastic depends on the input, i.e. different input values may have more noisy outputs than others. 
A common source of data uncertainty in recommender system is temporal variability, wherein the CTR of the same content will change over time. This variance is an inherent property of the content, and changes with the type of the content; 
for instance a trending fashion product will have large temporal variance. 
An optimistic exploration strategy (sec.~\ref{sec:optimistic_stragegy}) can exploit estimations of this variability by prioritizing content that has larger variance as it \textit{might} be trending at any given moment. 
This extends to non-temporal random variables that affect CTR and cannot be directly modeled.


We model data uncertainty by placing a distribution over the output of the model and learning it as a function of the input. 
To support that, we use a GMM with parameters ($\alpha_i$, $\mu_i$ and $\sigma_i$) to model our observation: 
$Y$:  

\begin{equation}
Y\sim \sum_i\alpha_i\mathcal{N}(\mu_i,\sigma_i^2)
\label{eq:mdn}
\end{equation}

\subsection{Measurement Uncertainty}
\label{sec:measurment_uncertainty}

Measurement uncertainty corresponds to the uncertainty of the observed CTR due to the measurement noise introduced by the binomial recommendation experiment. 
This type of uncertainty depends on the number of times $r$ a specific target $x=(t,c)$ pair was recommended, i.e. target $t$ was recommended in context $c$. 
In the previous section we saw how we can model data uncertainty by employing a GMM at the last layer of our network. 
However, since the observed CTR is affected by the measurement noise, the employed GMM gets polluted.
By modeling measurement noise separately we can remove this bias from the MDN.

Let $Y$,  $Y^*$ and $\epsilon$ be three random variables given $X=(t,c)$.
$Y$ corresponds to observed CTR, after recommending $(t,c)$ pair, $r$ times. 
$Y^*$ corresponds to the true/clean CTR without the measurement noise, i.e. the CTR if we had recommended $t$ infinite times in $c$.
$\epsilon$ corresponds to the binomial noise error distribution. 
\begin{equation}
\begin{gathered}
Y=Y^*+\epsilon, \quad \epsilon\sim \mathcal{N}(0,\sigma_\epsilon^2), \quad Y^*\sim \sum_i\alpha_i\mathcal{N}(\mu_i,\sigma_i^2)
\label{eq:y_y_e}
\end{gathered}
\end{equation}
%
%
%
%
We approximate measurement noise $\epsilon$ via a Gaussian model and model $Y^*$ with a GMM. 
%
%
For every $Y^*|X$ we enforce constant $\sigma_\epsilon=f(\mu, r)$, where $\mu$ is the expected value of $Y^*|X$.  
This way, $Y^* \indep \epsilon$ given $X$, as  $\sigma_\epsilon$ depends only on $r$ and $\mu$.
We can rewrite eq.~\ref{eq:y_y_e} and deconvolve data and measurement uncertainties. 
%
\begin{equation}
\begin{gathered}
Y \sim \sum_i\alpha_i\mathcal{N}(\mu_i,\sigma_i^2+\sigma_\epsilon^2)
\end{gathered}
\label{eq:sog}
\end{equation}
%
%
To this end, the DDN model described in sec.~\ref{sec:mdn} accounts for measurement uncertainty and predicts GMM's coefficients ($\alpha_i$, $\mu_i$ and $\sigma_i$),  from which we estimate the expected value and the standard deviation of $Y^*$.
%


\subsection{Model Uncertainty}
\label{sec:model_uncertainty}

Model uncertainty accounts for uncertainty in the model parameters. 
This corresponds to the ignorance of the model and depends on the data that the model was trained on. 
For example, if a recommendation system chooses to show mainly sports articles, 
future training datasets will contain mostly sports articles. As a result, the next 
trained model will have high model uncertainty for entertainment articles due to the lack of related content in the training dataset. 
This \textit{feedback loop} is common in recommendation systems;
trained model can only learn about areas of the features space that have been explored by previous models. 
This type of uncertainty, in contrast to data uncertainty, can be reduced if exploration is directed into areas of the feature space that were unexplored, making future models more robust to diverse types of recommendations.
We estimate model uncertainty using the Monte Carlo dropout method as a bayesian approximation introducted at~\cite{gal2016dropout}. 
Specifically, we train our DNN model using dropout and during inference we perform 
$T$ stochastic forward passes through the network, where $T$ is a tunable parameter. 
We collect $T$ estimations $\hat{y}_i$ and 
estimate model uncertainty as follows:
\begin{equation}
\begin{gathered}
\sigma = \sqrt{\frac{1}{T}\sum_{i=1}^T\hat{y}_i^2-(\frac{1}{T}\sum_{i=1}^T\hat{y_i})^2}
\end{gathered}
\label{eq:sog}
\end{equation}

\subsection{Optimistic strategy}
\label{sec:optimistic_stragegy}

Simple algorithms like $\epsilon$-greedy choose actions indiscriminately during exploration, with no specific preference for targets that have higher probability to be successful in exploitation. Uncertainty estimations allow to extend  $\epsilon$-greedy and employ the upper confidence bound (UCB) algorithm for better and adaptive exploration of new targets. For example, UCB will prioritize targets with titles that are composed of words that weren't previously recommended (via model uncertainty) and targets that have a larger variability in the CTR of their features (via data uncertainty). 


Our marketplace is defined by a very high recommendation turnover rate, with new content being uploaded every day and old one becoming obsolete. We allocate $\epsilon$ percent of our recommendation traffic to UCB; We estimate both the mean payoff $\mu^t$  and the standard deviation $\sigma^t$ of each target $t$ and select the target that achieves the highest score 
$A$ 
where $a$ is a tunable parameter.
\begin{equation}
\begin{gathered}
A=\arg\max_t(\mu^t+a\cdot\sigma^t)
\end{gathered}
\label{eq:ucb}
\end{equation}
%
%

\section{Evaluation}

This section contains two sets of results. First, we evaluate the effect of DDN modeling and of the various types of uncertainties, showing intuitive examples. Next, we show the impact of integrating DDN into Taboola's online recommendation engine. 

\subsection{Uncertainty estimations}

\begin{figure}[t]
\centering
\includegraphics[scale=0.45]{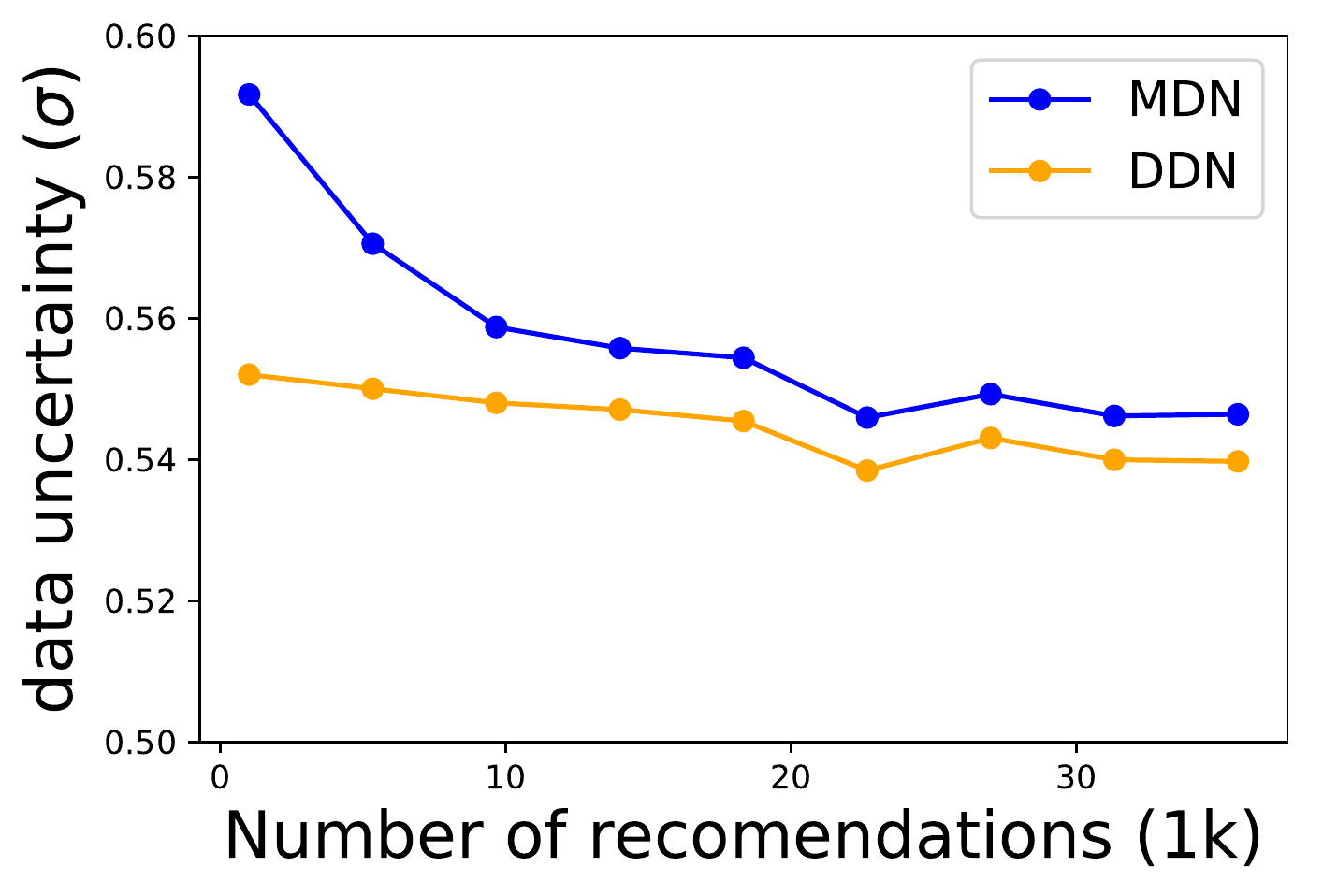}
\caption{Comparison between MDN and DDN models showing the effect of explicitly modeling measurement noise on data uncertainty.}
\label{fig:mdn_vs_ddn}
\end{figure}

\begin{figure}[t]
\centering
\includegraphics[scale=0.45]{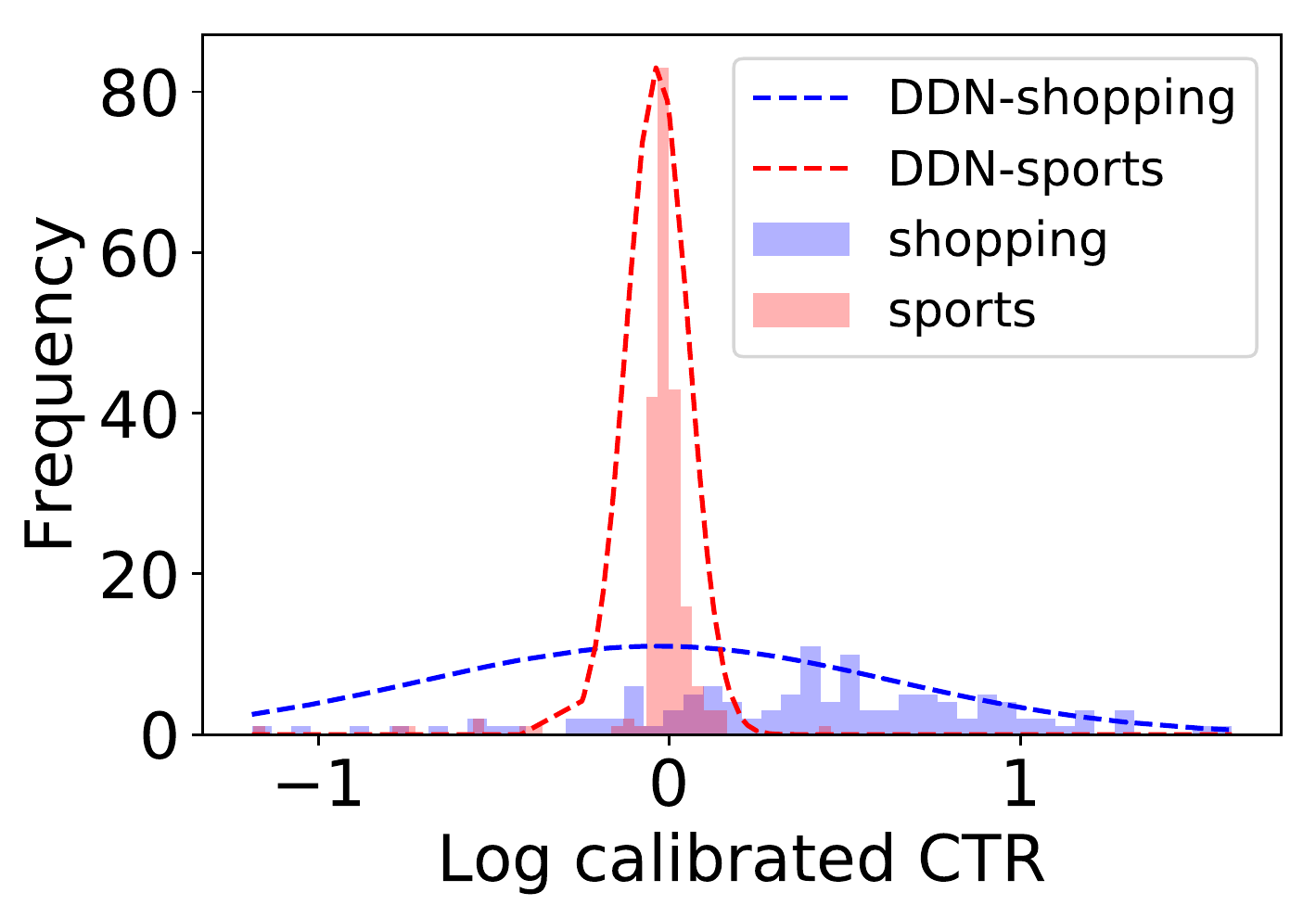}
\caption{Log calibrated CTR distribution for two groups of targets (shopping and sports) accompanied with DDN's data uncertainty estimation for a random target from each group.}
\label{fig:data_uncertainty}
\end{figure}

\begin{table*}[t]
	\begin{tabular}{ | l | l | }
	\hline
	shopping &sports \\	
	\hline
	\hline
	TOMS For \$35 - 41\% Off &Benfica vs Manchester United Betting\\
	Nike For \$72 - 40\% Off &It's The Only Way To Watch The Premier League\\
	Magnaflow Performance Mufflers From \$68.91 &LIVE: Arsenal vs Norwich City\\
	Sun Dolphin Mackinaw 15.6' Square Back Canoe &Premier League Castoffs  Starting Over at Age 11\\
	Brooks For \$65 - 35\% Off &Real Madrid Held to Draw With Tottenham\\
	ASICS For \$50 - 29\% Off &Rush for a 32/32 Score. NFL Team + City Match\\
	\hline
	\end{tabular}
	\caption{Indicative targets for the two selected groups (shopping and sports)}
	\label{table:samples_sports_shopping}
\end{table*}

\label{sec:uncertainty_examples}
For the results that follow in this subsection we have trained our models employing only title as feature vector for the target making the results human interpretable. 
%
%
In Fig.~\ref{fig:mdn_vs_ddn} we show the mean data uncertainty after bucketizing targets according to the number of times $r$ they have been recommended. 
We observe that the data uncertainty of the MDN model depends on $r$, i.e. low $r$ leads to high data uncertainty. 
This is an undesirable correlation; previously trained models chose to show more times recommendations from specific areas of the feature space, leading to reduced measurement uncertainty in the training set for those examples. 
MDN doesn't account for measurement noise explicitly, 
which causes a pollution of data uncertainty estimates. 
In contrast, DDN accounts for measurement uncertainty explicitly with the Gaussian model and is able both to reduce the predicted uncertainty and the aforementioned correlation significantly. 
This highlights the benefit of decorrelating measurement noise from data uncertainty (see sec.~\ref{sec:measurment_uncertainty}). 


In Fig.~\ref{fig:data_uncertainty} and Table~\ref{table:samples_sports_shopping} we study the nature of data uncertainty in the context of recommender systems. We first selected two groups of targets related to shopping and sports where intra-group targets are semantically close (see  Table~\ref{table:samples_sports_shopping}).
%
Further, we depict the CTR histogram of the two groups together with the distribution induced by the DDN prediction for one randomly selected target from each group (Fig.~\ref{fig:data_uncertainty}). We observe that the shopping group has large variability in CTR,  due to the fact that although all targets refer to shopping, the specific product that is being advertised highly affects CTR. 
This is in contrast to the sport group in which all sport related targets have relatively consistent CTR. 
We observe that the DDN model is able to capture and model this target-specific variability in the CTR 
and thus have the ability to exploit it in the optimistic strategy. 

\begin{figure}[t]
\centering
\includegraphics[scale=0.45]{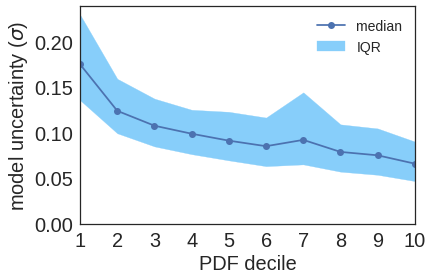}
\caption{DDN model uncertainty estimation after bucketizing targets according to their PDF values calculated using KDE over the training set. Low PDF value corresponds to big semantic distance from training set i.e. unexplored areas of the feature space.}
\label{fig:model_uncertainty_kde}
\end{figure}

As discussed in sec.~\ref{sec:model_uncertainty}, model uncertainty should capture what the model doesn't know. 
In order to validate this, we perform Kernel Density Estimation (KDE) over the targets' title feature representation in the training set,   
enabling us to quantify the semantic distance for each target from the training set. 
This way, targets located far away from the training examples i.e. belong to areas of the feature space which were less explored will have low Probability Distribution Function (PDF) value. 
In Fig.~\ref{fig:model_uncertainty_kde} we depict model uncertainty estimations for the DDN model after bucketizing targets in the validation set according to their PDF value relative to the training set. 
We observe that model uncertainty, is anti-correlated to the PDF value of the targets, 
indicating that DDN model indeed estimates high model uncertainty in less explored areas of the features space, which is  
a desirable behaviour in recommender systems. 
%

\begin{figure}[t]
\centering
\includegraphics[scale=0.45]{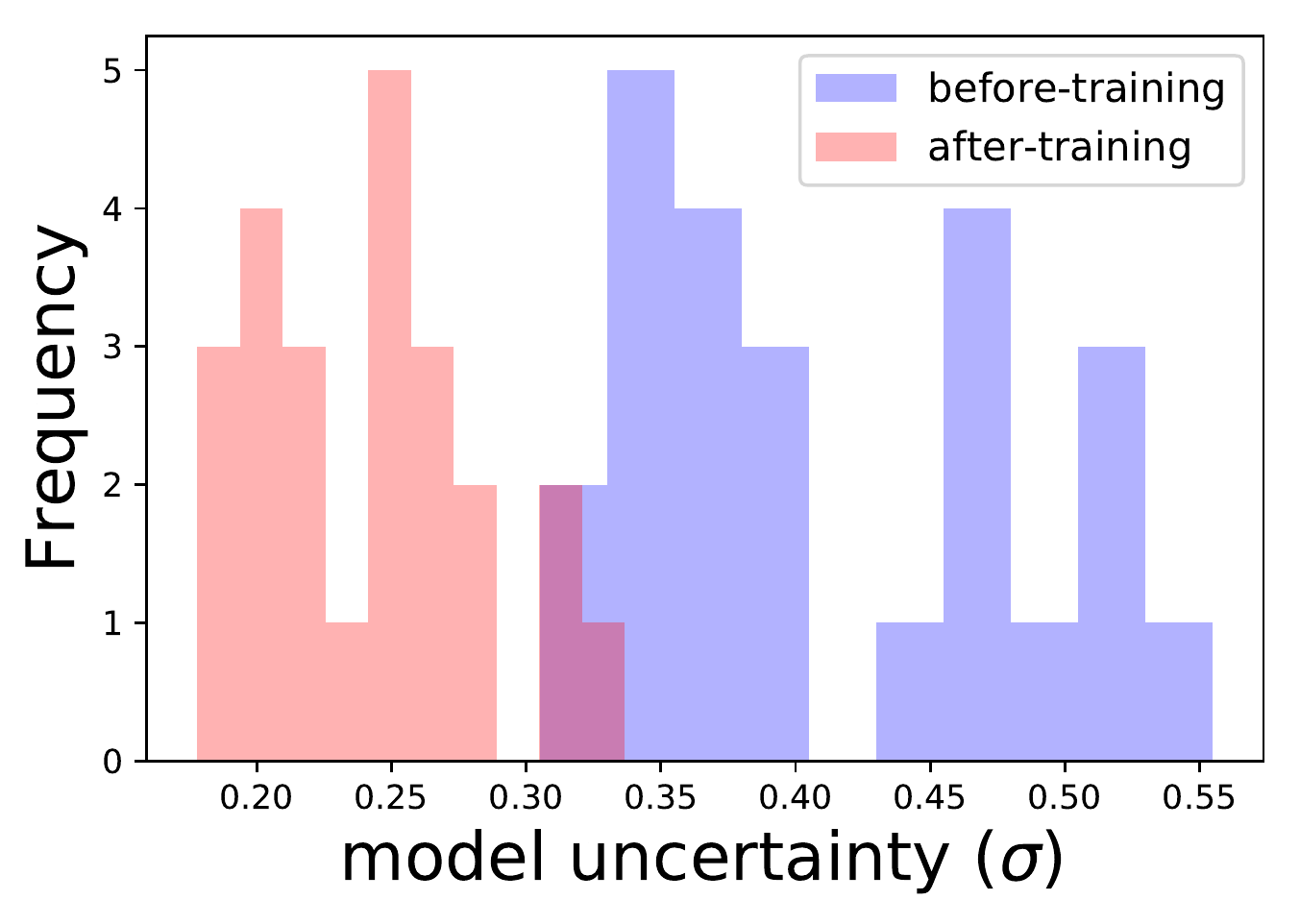}
\caption{Distribution of DDN model uncertainty estimation on selected group of targets, before and after employing during training the target "BMW X5".}
\label{fig:model_uncertainty}
\end{figure}

Another interesting observation is depicted in Fig.~\ref{fig:model_uncertainty} and Table~\ref{tab:model_uncertainty}, in which we show how model uncertainty is being affected while adding to the training set targets from unexplored areas of the feature space. 
We first selected a group of targets related to car advertisement with low PDF values 
and high model uncertainty.  
We then added one target from the group ("BMW X5") to the training set and retrained the model. 
We observe a reduction in the estimated model uncertainty of the group, indicating that pro-actively exploring targets with high model uncertainty can indeed lead to model uncertainty reduction. 


\begin{table}[t]
	\begin{tabular}{ | l | }
	\hline
	car related targets\\	
	\hline
	\hline
	2011 BMW M3 \\
	2005 Jaguar XK-Series XK8 Roadster \\
	2017 BMW X5 \\
	Mazda MX-5 Miata\\
	BMW X6 \\
	Find a BMW X5 Near You!\\
	The Fastest Car BMW i8\\
	\hline
	\end{tabular}
	\caption{Indicative targets for the selected group.}
	\label{tab:model_uncertainty}
\end{table}

\subsection{Performance evaluation}
\label{sec:online_experimentation}

\emph{Data:} We use the browsed website (i.e. $publisher$) as the user context for the following experiments. In all of the experiments we used three months of historical data for training, containing $\sim$10M records of target-publisher pairs. The dataset contains $\sim$1M unique targets and $\sim$10K unique publishers. 
%
Every offline experiment has been run on multiple time slots to validate that the results were statistically significant.

\emph{Models:}  In all models we performed an extension of the $\epsilon-greedy$ algorithm, where we allocate $\epsilon$ percent of the recommendation traffic to targets that have not been heavily exploited previously by the recommendation algorithm.

%
%

1. \textbf{REG} corresponds to our deep model described in sec.~\ref{sec:mdn}, where the output is just the predicted  CTR scalar employing MSE as loss as opposed to a GMM. 

2. \textbf{MDN} is similar to REG, with the GMM layer and the  
use of the optimistic strategy introduced in sec.~\ref{sec:optimistic_stragegy}. 


3. \textbf{DDN} is similar to MDN, with the measurement uncertainty modeling and optimistic strategy secs.~\ref{sec:measurment_uncertainty},~\ref{sec:optimistic_stragegy} .


In order to have a fair comparison, we tuned the hyper-parameters (e.g. embedding sizes, number of layers, number of mixtures) for each model separately; we performed thousands of iterations of random search, and chose the parameters that yielded the best results. We have found that this hyper-parameter tuning procedure was crucial in order to get the best possible results from our models, both offline and online. 

\emph{Metrics and evaluation:}
we use Mean Squared Error (MSE) for offline evaluation of our models. Due to the dynamic nature of online recommendations it is essential that we evaluate our models online within an A/B testing framework, by measuring the average RPM of models across different publishers. 
In addition, we utilize an online throughput metric which aims to capture the effectiveness of the exploration module; this metric counts the number of new targets that were discovered by the exploration mechanism at a certain given day by being shown significantly for the first time. 
We expect that exploration models which are better at exploring the feature space will learn to recommend more from this pool of new targets. Similarly, we have a metric for advertisers throughput. 
In addition to RPM dynamics, maintaining high throughput levels is essential to ensure advertiser satisfaction levels.

%
%


\begin{table}[t]
\centering
	\begin{tabular}{ | c | c | c | c | c |}
	\hline
	Model&REG&MDN&DDN\\
	\hline
	RPM lift&0\%& 1.2\%& 2.9\%\\
	\hline
	\end{tabular}
\caption{A comparison of the online RPM lift between the different models}
\label{tab:model_comparison}
\end{table}

\begin{table}[t]
	\begin{tabular}{ | c | c | c | c | }
	\hline
	Dataset &MDN&DDN&Improvement\\	
	\hline
	D1 &0.2681&0.25368&5.3\%\\
	\hline
	D2 & 0.25046&0.24367&2.7 \% \\
	\hline
	\end{tabular}
	\caption{Relative improvement in the MSE between MDN and DDN when trained over two datasets that differ by the amount of measurement noise.}
	\label{tab:varying_wrecs}
 \end{table}

\subsubsection{Experimental results}

%
%

\textbf{Model comparison:} in Table~\ref{tab:model_comparison} we compare the three different models discussed previously in terms of online RPM. We observe that both  \textbf{MDN} and  \textbf{DDN} outperform  \textbf{REG} by 1.2\% and 2.9\% respectively. 
Although the improvements may seem small numerically, they have a large product impact as they translate to significantly higher revenue. In addition, it's  noteworthy that \textbf{REG} is a highly optimized and tuned model which is our current state-of-the-art model making it a very competitive baseline to win. 
These results verify once again that the loss attenuation achieved during training has enabled the model to converge to better parameters,  generalizing  better to unseen examples.
Furthermore we observe that \textbf{DDN} outperforms \textbf{MDN} by 1.7\%, indicating that deconvolving measurement noise from the data uncertainty leads to further gains. 

\textbf{Measurement noise:} in Table~\ref{tab:varying_wrecs} we compare the MDN and DDN models by training them on two different datasets, D1 and D2. D1 differs from D2 by the amount of noise in the training samples; D1 contains noisy data points with relatively small amount of empirical data, while D2 contains examples with higher empirical statistical significance. We observe that DDN improves on MDN performance by 2.7\% when using D1 for training, and by 5.3\% when using D2. This validates that integrating measurement noise into our modeling is crucial when the training data contains very noisy samples, by attenuating the impact of measurement noise on the loss function.  (see sec.~\ref{sec:mdn})

\begin{table}[t]
\begin{center}
	\begin{tabular}{| c | c | c | c | c |}
	\hline
	$a$ &0&0.5&1&1.5\\
	\hline
	\hline
	RPM lift &0\%& -0.05\%&-0.2\%&-0.3\%\\
	Target throughput lift &0\%&6.5\%&9.1\%&11.7\%\\
	Advertiser throughput lift &0\%&2.1\%&3.7\%&5.1\%\\
	\hline
	\end{tabular}
		\caption{RPM lift, targets and advertisers throughput as a function of different values of $a$.}
	\label{tab:rpm_throughput}
\end{center}
\end{table}

\textbf{RPM lift vs. targets throughput:}  we analyzed the effect of the parameter $a$ found in eq.~\ref{eq:ucb} by employing data uncertainty. From a theoretical standpoint, increasing this value is supposed to prioritize higher information gain at the expense of RPM, by choosing targets with higher uncertainty. This trade-off is worthwhile in the long term. In Table~\ref{tab:rpm_throughput} we observe that there is an inverse correlation between RPM and throughput which is triggered by different values of $a$, with targets and advertisers throughput increasing by 11.7\% and 5.1\% respectively when setting $a=1.5$.  Choosing the right trade-off depends on the application and the business KPIs. For our case we chose $a = 0.5$, resulting in a good throughput gain with a small RPM cost.

\section{Conclusions}

We have introduced Deep Density Network (DDN), a hybrid unified DNN model that estimates uncertainty. DDN is able to model non-linearities and capture complex target-context relations, incorporating higher level representations of data sources such as contextual and textual input. We presented the various types of uncertainties that might arise in recommendation systems, and investigated the effect of integrating them into the recommendation model. We have shown the added value of using DNN in a multi-arm bandit setting, yielding an adaptive selection strategy that balances \textit{exploitation} and \textit{exploration} and maximizes the long term reward. We presented results validating DDN's improved noise handling capabilities, leading to 5.3\% improvement on a noisy dataset.  Furthermore,  DDN outperformed both REG and MDN models in online  experiments, leading to RPM improvements of 2.9\% and 1.7\% respectively. 
Finally, by employing DDN's 
uncertainty estimation and optimistic strategy, we improved our exploration strategy, depicting 6.5\% and 2.1\% increase of targets and advertisers throughput respectively with only 0.05\% RPM decrease. 


\bibliographystyle{ACM-Reference-Format}
\bibliography{mdn_kdd2018}


\begin{thebibliography}{18}


\ifx \showCODEN    \undefined \def \showCODEN     #1{\unskip}     \fi
\ifx \showDOI      \undefined \def \showDOI       #1{#1}\fi
\ifx \showISBNx    \undefined \def \showISBNx     #1{\unskip}     \fi
\ifx \showISBNxiii \undefined \def \showISBNxiii  #1{\unskip}     \fi
\ifx \showISSN     \undefined \def \showISSN      #1{\unskip}     \fi
\ifx \showLCCN     \undefined \def \showLCCN      #1{\unskip}     \fi
\ifx \shownote     \undefined \def \shownote      #1{#1}          \fi
\ifx \showarticletitle \undefined \def \showarticletitle #1{#1}   \fi
\ifx \showURL      \undefined \def \showURL       {\relax}        \fi
\providecommand\bibfield[2]{#2}
\providecommand\bibinfo[2]{#2}
\providecommand\natexlab[1]{#1}
\providecommand\showeprint[2][]{arXiv:#2}

\bibitem[\protect\citeauthoryear{Auer, Cesa-Bianchi, and Fischer}{Auer
  et~al\mbox{.}}{2002}]%
        {auer2002finite}
\bibfield{author}{\bibinfo{person}{Peter Auer}, \bibinfo{person}{Nicolo
  Cesa-Bianchi}, {and} \bibinfo{person}{Paul Fischer}.}
  \bibinfo{year}{2002}\natexlab{}.
\newblock \showarticletitle{Finite-time analysis of the multiarmed bandit
  problem}.
\newblock \bibinfo{journal}{\emph{Machine learning}} \bibinfo{volume}{47},
  \bibinfo{number}{2-3} (\bibinfo{year}{2002}), \bibinfo{pages}{235--256}.
\newblock


\bibitem[\protect\citeauthoryear{Bishop}{Bishop}{1994}]%
        {bishop1994mixture}
\bibfield{author}{\bibinfo{person}{Christopher~M Bishop}.}
  \bibinfo{year}{1994}\natexlab{}.
\newblock \showarticletitle{Mixture density networks}.
\newblock \bibinfo{journal}{\emph{Technical report}} (\bibinfo{year}{1994}).
\newblock


\bibitem[\protect\citeauthoryear{Blundell, Cornebise, Kavukcuoglu, and
  Wierstra}{Blundell et~al\mbox{.}}{2015}]%
        {blundell2015weight}
\bibfield{author}{\bibinfo{person}{C. Blundell}, \bibinfo{person}{J.
  Cornebise}, \bibinfo{person}{K. Kavukcuoglu}, {and} \bibinfo{person}{D.
  Wierstra}.} \bibinfo{year}{2015}\natexlab{}.
\newblock \showarticletitle{Weight uncertainty in neural networks}.
\newblock \bibinfo{journal}{\emph{arXiv preprint arXiv:1505.05424}}
  (\bibinfo{year}{2015}).
\newblock


\bibitem[\protect\citeauthoryear{Chamiel, Golan, A., Sinai, and
  Pilberg}{Chamiel et~al\mbox{.}}{2013}]%
        {TabPatent}
\bibfield{author}{\bibinfo{person}{G. Chamiel}, \bibinfo{person}{L. Golan},
  \bibinfo{person}{Rubin A.}, \bibinfo{person}{M. Sinai, A.~Salomon}, {and}
  \bibinfo{person}{A. Pilberg}.} \bibinfo{year}{2013}\natexlab{}.
\newblock \showarticletitle{Click through rate estimation in varying display
  situations}.
\newblock \bibinfo{journal}{\emph{United States Patent Application
  Publication}} (\bibinfo{year}{2013}).
\newblock


\bibitem[\protect\citeauthoryear{Cheng, Koc, Harmsen, Shaked, Chandra, Aradhye,
  Anderson, Corrado, Chai, Ispir, et~al\mbox{.}}{Cheng et~al\mbox{.}}{2016}]%
        {cheng2016wide}
\bibfield{author}{\bibinfo{person}{Heng-Tze Cheng}, \bibinfo{person}{Levent
  Koc}, \bibinfo{person}{Jeremiah Harmsen}, \bibinfo{person}{Tal Shaked},
  \bibinfo{person}{Tushar Chandra}, \bibinfo{person}{Hrishi Aradhye},
  \bibinfo{person}{Glen Anderson}, \bibinfo{person}{Greg Corrado},
  \bibinfo{person}{Wei Chai}, \bibinfo{person}{Mustafa Ispir}, {et~al\mbox{.}}}
  \bibinfo{year}{2016}\natexlab{}.
\newblock \showarticletitle{Wide \& deep learning for recommender systems}. In
  \bibinfo{booktitle}{\emph{Proceedings of the 1st Workshop on Deep Learning
  for Recommender Systems}}. ACM, \bibinfo{pages}{7--10}.
\newblock


\bibitem[\protect\citeauthoryear{Fr{\'e}nay and Verleysen}{Fr{\'e}nay and
  Verleysen}{2014}]%
        {frenay2014classification}
\bibfield{author}{\bibinfo{person}{Beno{\^\i}t Fr{\'e}nay} {and}
  \bibinfo{person}{Michel Verleysen}.} \bibinfo{year}{2014}\natexlab{}.
\newblock \showarticletitle{Classification in the presence of label noise: a
  survey}.
\newblock \bibinfo{journal}{\emph{IEEE transactions on neural networks and
  learning systems}} \bibinfo{volume}{25}, \bibinfo{number}{5}
  (\bibinfo{year}{2014}), \bibinfo{pages}{845--869}.
\newblock


\bibitem[\protect\citeauthoryear{Gal and Ghahramani}{Gal and
  Ghahramani}{2016}]%
        {gal2016dropout}
\bibfield{author}{\bibinfo{person}{Y. Gal} {and} \bibinfo{person}{Z.
  Ghahramani}.} \bibinfo{year}{2016}\natexlab{}.
\newblock \showarticletitle{Dropout as a Bayesian approximation: Representing
  model uncertainty in deep learning}. In \bibinfo{booktitle}{\emph{Intl' Conf.
  on machine learning}}. \bibinfo{pages}{1050--1059}.
\newblock


\bibitem[\protect\citeauthoryear{Goldberger and Ben-Reuven}{Goldberger and
  Ben-Reuven}{2017}]%
        {goldberger2016training}
\bibfield{author}{\bibinfo{person}{Jacob Goldberger} {and}
  \bibinfo{person}{Ehud Ben-Reuven}.} \bibinfo{year}{2017}\natexlab{}.
\newblock \showarticletitle{Training deep neural-networks using a noise
  adaptation layer}. In \bibinfo{booktitle}{\emph{ICLR}}.
\newblock


\bibitem[\protect\citeauthoryear{Kendall and Gal}{Kendall and Gal}{2017}]%
        {kendall2017uncertainties}
\bibfield{author}{\bibinfo{person}{A. Kendall} {and} \bibinfo{person}{Y. Gal}.}
  \bibinfo{year}{2017}\natexlab{}.
\newblock \showarticletitle{What Uncertainties Do We Need in Bayesian Deep
  Learning for Computer Vision?}
\newblock \bibinfo{journal}{\emph{arXiv preprint arXiv:1703.04977}}
  (\bibinfo{year}{2017}).
\newblock


\bibitem[\protect\citeauthoryear{Kingma and Welling}{Kingma and
  Welling}{2013}]%
        {kingma2013auto}
\bibfield{author}{\bibinfo{person}{Diederik~P Kingma} {and}
  \bibinfo{person}{Max Welling}.} \bibinfo{year}{2013}\natexlab{}.
\newblock \showarticletitle{Auto-encoding variational bayes}.
\newblock \bibinfo{journal}{\emph{arXiv preprint arXiv:1312.6114}}
  (\bibinfo{year}{2013}).
\newblock


\bibitem[\protect\citeauthoryear{Li, Chu, Langford, and Schapire}{Li
  et~al\mbox{.}}{2010}]%
        {li2010contextual}
\bibfield{author}{\bibinfo{person}{Lihong Li}, \bibinfo{person}{Wei Chu},
  \bibinfo{person}{John Langford}, {and} \bibinfo{person}{Robert~E Schapire}.}
  \bibinfo{year}{2010}\natexlab{}.
\newblock \showarticletitle{A contextual-bandit approach to personalized news
  article recommendation}. In \bibinfo{booktitle}{\emph{Proceedings of the 19th
  international conference on World wide web}}. ACM, \bibinfo{pages}{661--670}.
\newblock


\bibitem[\protect\citeauthoryear{Linden, Smith, and York}{Linden
  et~al\mbox{.}}{2003}]%
        {linden2003amazon}
\bibfield{author}{\bibinfo{person}{Greg Linden}, \bibinfo{person}{Brent Smith},
  {and} \bibinfo{person}{Jeremy York}.} \bibinfo{year}{2003}\natexlab{}.
\newblock \showarticletitle{Amazon. com recommendations: Item-to-item
  collaborative filtering}.
\newblock \bibinfo{journal}{\emph{IEEE Internet computing}}
  \bibinfo{volume}{7}, \bibinfo{number}{1} (\bibinfo{year}{2003}),
  \bibinfo{pages}{76--80}.
\newblock


\bibitem[\protect\citeauthoryear{Mnih and Salakhutdinov}{Mnih and
  Salakhutdinov}{2008}]%
        {mnih2008probabilistic}
\bibfield{author}{\bibinfo{person}{Andriy Mnih} {and} \bibinfo{person}{Ruslan~R
  Salakhutdinov}.} \bibinfo{year}{2008}\natexlab{}.
\newblock \showarticletitle{Probabilistic matrix factorization}. In
  \bibinfo{booktitle}{\emph{Advances in neural information processing
  systems}}. \bibinfo{pages}{1257--1264}.
\newblock


\bibitem[\protect\citeauthoryear{Mnih and Hinton}{Mnih and Hinton}{2012}]%
        {mnih2012learning}
\bibfield{author}{\bibinfo{person}{V. Mnih} {and} \bibinfo{person}{G. Hinton}.}
  \bibinfo{year}{2012}\natexlab{}.
\newblock \showarticletitle{Learning to label aerial images from noisy data}.
  In \bibinfo{booktitle}{\emph{Proc. of the 29th Intl Conf. on Machine Learning
  (ICML-12)}}. \bibinfo{pages}{567--574}.
\newblock


\bibitem[\protect\citeauthoryear{Neal}{Neal}{2012}]%
        {neal2012bayesian}
\bibfield{author}{\bibinfo{person}{Radford~M Neal}.}
  \bibinfo{year}{2012}\natexlab{}.
\newblock \bibinfo{booktitle}{\emph{Bayesian learning for neural networks}}.
  Vol.~\bibinfo{volume}{118}.
\newblock \bibinfo{publisher}{Springer Science \& Business Media}.
\newblock


\bibitem[\protect\citeauthoryear{Rezende, Mohamed, and Wierstra}{Rezende
  et~al\mbox{.}}{2014}]%
        {rezende2014stochastic}
\bibfield{author}{\bibinfo{person}{Danilo~Jimenez Rezende},
  \bibinfo{person}{Shakir Mohamed}, {and} \bibinfo{person}{Daan Wierstra}.}
  \bibinfo{year}{2014}\natexlab{}.
\newblock \showarticletitle{Stochastic backpropagation and approximate
  inference in deep generative models}.
\newblock \bibinfo{journal}{\emph{arXiv preprint arXiv:1401.4082}}
  (\bibinfo{year}{2014}).
\newblock


\bibitem[\protect\citeauthoryear{Thompson}{Thompson}{1933}]%
        {thompson1933likelihood}
\bibfield{author}{\bibinfo{person}{W.~R Thompson}.}
  \bibinfo{year}{1933}\natexlab{}.
\newblock \showarticletitle{On the likelihood that one unknown probability
  exceeds another in view of the evidence of two samples}.
\newblock \bibinfo{journal}{\emph{Biometrika}} \bibinfo{volume}{25},
  \bibinfo{number}{3/4} (\bibinfo{year}{1933}).
\newblock


\bibitem[\protect\citeauthoryear{Zhu and Laptev}{Zhu and Laptev}{2017}]%
        {zhu2017deep}
\bibfield{author}{\bibinfo{person}{L. Zhu} {and} \bibinfo{person}{N. Laptev}.}
  \bibinfo{year}{2017}\natexlab{}.
\newblock \showarticletitle{Deep and Confident Prediction for Time Series at
  Uber}. In \bibinfo{booktitle}{\emph{Data Mining Wrkshp}}. IEEE,
  \bibinfo{pages}{103--110}.
\newblock


\end{thebibliography}

\end{document}